\documentclass{article}

\usepackage{graphicx}
\usepackage{wrapfig}

\usepackage{array}
\usepackage{multirow}
\usepackage{tabularx}
\usepackage[para,online,flushleft]{threeparttable}
\usepackage{makecell}

\usepackage{booktabs}
\usepackage{pifont}
\usepackage{lipsum}
\usepackage[numbers]{natbib}
\usepackage{booktabs}
\usepackage{siunitx}

\usepackage{url}

\def\BibTeX{{\rm B\kern-.05em{\sc i\kern-.025em b}\kern-.08emT\kern-.1667em\lower.7ex\hbox{E}\kern-.125emX}}

\begin{document}
\title{Threat Landscape for Smart Grid Systems}

\author{Christos-Minas Mathas, Konstantinos-Panagiotis Grammatikakis,\\ Costas Vassilakis, Nicholas Kolokotronis, Vasiliki-Georgia Bilali,\\ and Dimitris Kavallieros%
\thanks{C.-M. Mathas, K.-P. Grammatikakis, C. Vassilakis, and N. Kolokotronis are with the Department of Informatics and Telecommunications, University of the Peloponnese, Tripolis, Greece (e-mail \{mathas.ch.m,\,kpgram,\,costas,\,nkolok\}@uop.gr)%
\newline\indent%
V.-G. Bilali is with the Center for Security Studies -- KEMEA, Athens, Greeece (e-mail g.bilali@kemea-research.gr)%
\newline\indent%
D. Kavallieros is with the Center for Security Studies -- KEMEA, Athens, Greeece (e-mail d.kavallieros@kemea-research.gr) and the Department of Informatics and Telecommunications, University of the Peloponnese, Tripolis, Greece (e-mail d.kavallieros@uop.gr)%
}}

\date{}

This paper is a preprint; it has been accepted for publication in the 15th International Conference on Availability, Reliability and Security (ARES 2020), August 25--28, 2020, Virtual Event, Ireland (DOI: 10.1145/3407023.3409229).
\bigskip

\noindent{\bf ACM copyright notice}

\noindent%
Permission to make digital or hard copies of all or part of this work for personal or classroom use is granted without fee provided that copies are not made or distributed for profit or commercial advantage and that copies bear this notice and the full citation on the first page. Copyrights for components of this work owned by others than the author(s) must be honored. Abstracting with credit is permitted. To copy otherwise, or republish, to post on servers or to redistribute to lists, requires prior specific permission and/or a fee. Request permissions from \url{permissions@acm.org}.

\maketitle

\begin{abstract}
\noindent%
Smart Grids are energy delivery networks, constituting an evolution of power grids, in which a bidirectional flow between power providers and consumers is established. These flows support the transfer of electricity and information, in order to support automation actions in the context of  
the energy delivery network. Insofar, many smart grid implementations and implementation proposals have emerged, with varying degrees of feature delivery and sophistication. While smart grids offer many advantages, their distributed nature and information flow streams between energy producers and consumers enable the launching of a number of attacks against the smart grid infrastructure, where the related consequences may range from economic loss to complete failure of the smart grid. In this paper, we survey the threat landscape of smart grids, identifying threats that are specific to this infrastructure, providing an assessment of the severity of the consequences of each attack type, discerning features that can be utilized to detect attacks and listing methods that can be used to mitigate them.
\bigskip

\end{abstract}

\section{Introduction}
Smart Grids are energy delivery networks, constituting an evolution of power grids, in which a bidirectional flow between power providers and consumers is established. These flows support the transfer of electricity and information, in order to support automation actions in the context of  
the energy delivery network \cite{Fang2012}. The information flowing in smart grids is produced by IoT devices termed as \textit{smart meters }, and this information is used to support a multitude of applications, including billing automation; accurate estimation of electricity needs to both (a) regulate power production and (b) raise/lower the cost accordingly; reduce power outages; confront energy theft \cite{Kaplantzis2012}. These developments contribute to the minimization of operational losses and the conservation of energy, towards the realization of a more ``green" power production and distribution network.

\begin{figure}[h!]
  \centering
  \includegraphics[width=0.75\columnwidth]{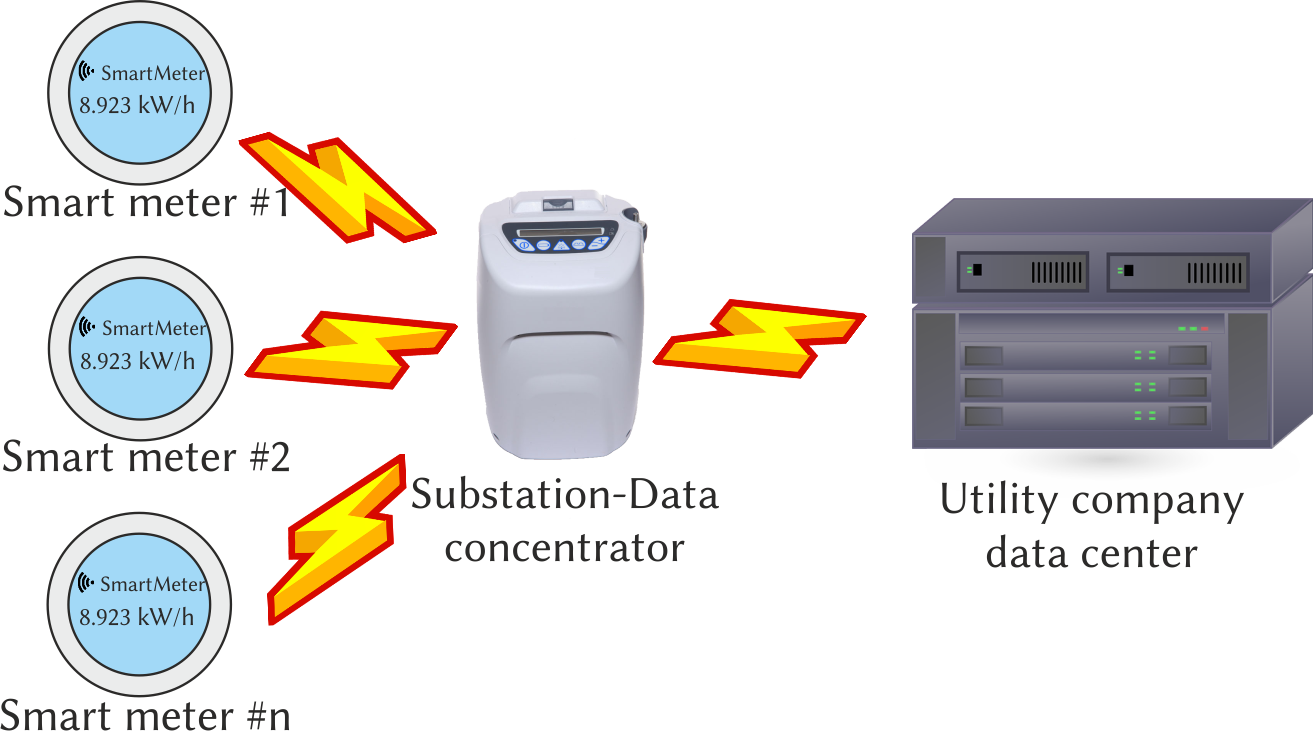}
  \caption{Smart grid metering and control system}
  \label{fig:smartGridNetwork}
\end{figure}

\begin{figure}[h!]
  \centering
  \includegraphics[width=0.75\columnwidth]{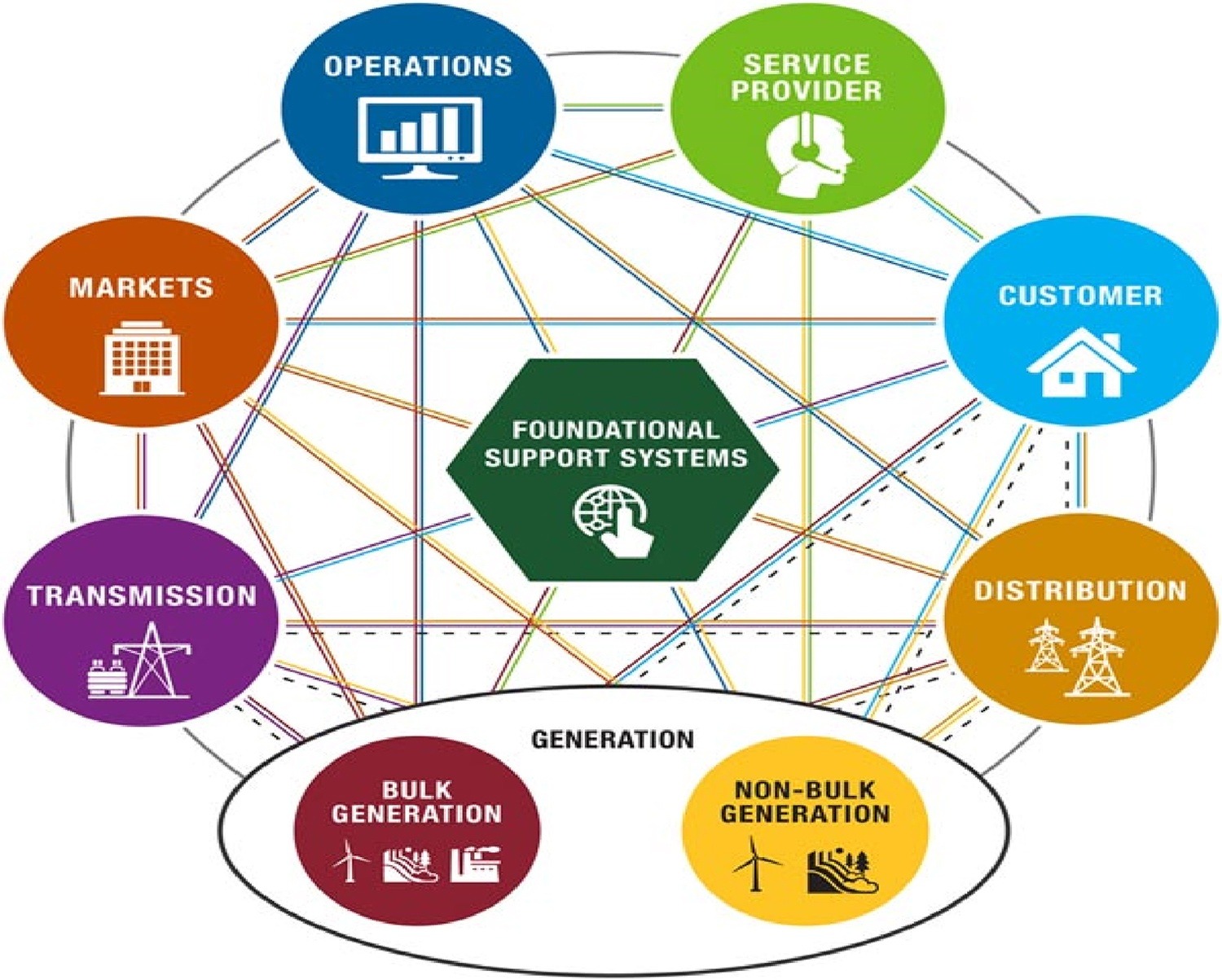}
  \caption{Smart grid environment (source: \cite{NIST2014})}
  \label{fig:smartgridarchitecture}
\end{figure}

A typical smart-grid metering and control system consists of a collection of meters that communicate with a substation/data-concentrator, realizing the home/installation area of the smart grid, uploading to the concentrator energy consumption data. The data concentrator of the home/installation area of the smart grid communicates with the utility company data center, sending to it detailed and aggregated energy consumption data. The communication among different smart grid entities is realized by high-speed wired or wireless links or a combination thereof, forming the smart grid network. A smart-grid metering and control system has a layered network structure through which it collects data and controls the delivery of electricity \cite{NIST2014,ENISA2016,Fan2013}. As noted in \cite{Fan2013}, the most commonly used smart meter-to-concentrator communication protocols include WiFi, Zigbee and power line carrier (PLC), while the communication between the concentrator and the utility company data center can be realized Wi-Fi, satellite, 4G-LTE, Wi-Max etc., or combination of these protocols and/or public internet lines. The smart grid metering and control system is depicted in Figure \ref {fig:smartGridNetwork}, while the overall smart grid environment, illustrating the smart grid's wider operational context, is depicted in Fig. \ref{fig:smartgridarchitecture} (source: \cite{NIST2014}). 

The assets comprising a smart grid are bound to face a number of threats, originating from a variety of threat agents and having diverse results and different impact levels \cite{Goel2015,Khelifa2015,NISTIR2014,Li2010,sanjab2016smart,Esmalifalak2013,Li2012}. A considerable class of threats concerns the physical dimension of the smart grid: equipment physically damaged due to natural phenomena (e.g. thunders, floods, earthquakes) or due to deliberate activities -including terrorist attacks or consumer attempts to evade charging; equipment can also be stolen or subject to physical wear-out, due to the passage of time and exposure to adverse environmental conditions. A second class of threats comprises cyber-attacks against components of the smart grid environment; in these types of attacks, smart grid network data or endpoint devices are manipulated to serve the purposes of attackers. In this paper, we will focus on the second class of threats only. In the context of these threats, cyber-attackers may span across a multitude of categories, ranging from individuals attempting to evade energy consumption charges, activists trying to disrupt specific aspects of the smart grid operation, misconducting energy providers targeting to the manipulation of energy market prices and therefore to the increase of their benefits, or professional hackers seeking to gain access to personal data that will be subsequently sold or used for the extortion of legal asset owners. The impact of cyber-attacks may include leakage of varying magnitudes of data, from which important information about individuals may be inferred; economic fraud involving reduced client billings or energy market price manipulation; instantaneous or prolonged smart grid malfunction. Smart grid malfunctions, especially prolonged ones, may have extremely severe consequences, such as environmental hazards/pollution, rendering of critical infrastructures, such as hospitals or security defenses, inoperable, suspension of economic activities and so forth. Considering the above, it is imperative that smart grid security officers are highly aware of potential attacks against the systems they protect, are able to prioritize attack mitigations, according to the key security goals of their organizations and possess the knowledge on how to detect and mitigate attacks. 

This paper contributes to the state of the art as follows:
\begin{itemize}
    \item it synthesizes content from existing reviews on the smart grid threat landscape;
    \item it covers newly added threats, taking into account that the smart grid threat landscape is constantly evolving
    \item it provides information on how attacks on smart grid infrastructure can be detected and mitigated.
\end{itemize}

This paper is structured as follows:  section \ref{sec:relWork} overviews related work in the domain of smart grid security, while section \ref {sec:keyGoals} presents the key security goals related to smart grid systems. Section \ref{sec:attackTypes} presents the attack types specific to smart grid systems and section \ref{sec:attackDetectionAndMitigation} lists the techniques and mechanisms that can be employed to detect and mitigate attacks. Finally, section \ref{sec:conclusions} concludes the paper.

\section{Related work}
\label{sec:relWork}

In this section we present the research related to smart grid threats and countermeasures. We begin with three works that discuss the threat landscape as well as existing solutions specifically for the smart grid domain.

The work in \cite{ENISA2013}, discusses the smart grid threat landscape and provides a good practice guide, amongst other matters. It focuses mainly on cyber-security threats, while also mentioning some threats to physical assets that are critical to the operation of the smart grid infrastructure. The good practice guide provided categorizes the security measures to those pertaining to “IT Systems and Logical Networks” and “Supply chain” and provides an extensive list of high-level best practices. 

In \cite{Ghansah2010}, an extensive report on the smart grid cyber security threats, vulnerabilities and risks is provided, which covers the security issues that reside in various domains of the smart grid infrastructure, such as customer domain, demand response, SCADA, and more. Furthermore, the work discusses the best practices to be followed for each domain.

In \cite{Fang2012}, the authors categorize the attacks against smart grids under the CIA triad, but the range of attacks discussed is limited. Additionally, they provide a survey of the cyber-security solutions proposed by the industry and the academia, while discussing some of the solutions in short.

The threats and corresponding mitigations pertaining to Critical Information Infrastructures (CII), a superset of smart grids are discussed in \cite{Europol2016} and \cite{Simon2017}.
In \cite{Europol2016}, a special chapter is dedicated in discussing the most prominent attacks on critical infrastructure. The report identifies the ``key threats" against CII, including but not limited to, attacks against the infrastructure grid, providing specific examples, and the shift of malicious actors to social engineering practices. The recommendations provided in this chapter, do not focus on the threats mentioned, instead mention the need for law enforcement and regulations.
In \cite{Simon2017}, the cyber threats to critical infrastructure are discussed, by incorporating the factor of the Industrial Internet of Things (IIoT). However, threats and recommendations are only discussed at a high-level. The ever-evolving threat landscape of smart grids is portrayed in a number of additional surveys, including \cite{Goel2015,Kaplantzis2012,Khelifa2015,Li2012,NIST2014}. It is worth noting that for a number of attacks no actual details have been released, due to security concerns (e.g. \cite{Hacket2017}).

Regarding architectures that have been proposed in the literature, aiming at securing critical information infrastructures, \cite{Stouffer2015} offers an extensive guide to industrial control systems (ICS) security, identifying the risks present in ICS deployments and providing detailed recommended security countermeasures to mitigate them, forming a secure ICS architecture model.
In \cite{Obregon2015}, a reference ICS architecture is defined which is used as the basis for presenting secure architectural patterns in four ICS domains, namely: access control, log management, network security, and remote access.
In \cite{Verssimo2006} and 
\cite{Verssimo2008}, the authors propose a secure architecture for critical information infrastructures, incorporating a set of techniques and algorithms in order to achieve fault tolerance and attack resiliency in an automatic way.

\section{Key security goals for smart grids}
\label{sec:keyGoals}

Smart grids are (parts of) critical information infrastructures, and in this respect their operation affects a number of assets, some of which have a very high impact in everyday life and/or specific operations segments. The information recorded and transmitted by smart meters is of high importance, since energy providers use this to (i) regulate energy production and energy flow within the distribution network and (ii) charge consumers for the energy they use. Energy consumers may also use this data for optimizing their energy use. It must also be noted that energy consumption data can be analysed to infer patterns, which can disclose life schedules, personal habits and events, an approach termed as non-intrusive load monitoring (NLM) \cite{Wang2012}; clearly  NLM adversely affects user privacy \cite{Lisovich2010}. According to the literature \cite{Bellias2016,Anderson2014,EC2018,Vassilakis2018} smart grids should meet a number of key security goals as follows:

\begin{enumerate}
\item \textit{Safety}, i.e. the system or the devices should operate without causing any risk to technological services, public services, humans or even to the environment. In the context of smart grids, safety refers not only to the fact that the smart grid infrastructure should not cause any harm, but also no harm should be caused by the malfunction of services that rely on smart grids infrastructure and data.
\item \textit{Security}, i.e. the protection of the system from unintended or unauthorized access, change/disruption or destruction (e.g. malware, remote attacks)
\item \textit{Reliability}, i.e. the ability of the smart to perform its required functions under stated conditions; this spans across smart grid components, e.g. smart meters.
\item \textit{Resilience}, i.e. the ability to withstand and operate as normal as possible while being under major disruption. 
\item \textit{Privacy}, i.e. data should be available only to the owner of the data and expressly and legally authorized parties. Smart meter collects data that could be analysed in order to infer information that could severely threaten privacy and also have additional consequences; such information include the time people are at home, living habits and so forth.
\item \textit{Accuracy}, meaning that the system should correctly calculate energy consumption and support the efficient distribution of information.
\item \textit{Availability of resources} at any given time. Both the energy provider as well as the consumer must have access to the respective information e.g. billing information, control messages.
\item \textit{Integrity}, i.e. the ability of the system to prevent any changes of the collected data as well as control commands.
\end{enumerate}

Taking the above into account, we can conclude that in the context of smart meters, protection from harm, protection of the environment, resilience, operations reliability \& continuity, as well as maintenance of data integrity and confidentiality/privacy are the key high-level security goals. Any threat that jeopardizes the aforementioned goals should be assessed and treated accordingly.

\section{Main attack types against smart grids}
\label{sec:attackTypes}

The current literature reports numerous types of attacks on smart grids; each attack targets one or more of the fundamental information security dimensions, i.e. confidentiality, integrity and availability. The attack goals are contextualized in the smart grid domain as follows:

\begin{itemize}

\item \textit{confidentiality}: attackers attempt to acquire data flowing within the smart grid network. Data can be acquired at transmission time or at the endpoint smart-grid nodes. It is worth noting that data can be also exfiltrated from information systems related to the smart grid, e.g. from power company billing systems. These information systems are typically targeted by typical attack methods, and are considered to be outside the scope of this work.

\item \textit{integrity}: attackers aim at altering the data flowing  within the smart grid network. Unauthorized injection of data or modification of legitimate data are the most common attack types of this class.

\item \textit{availability}: these attacks attempt to disrupt the operation of the smart grid, delaying, corrupting or interrupting network communications or the operation of endpoint devices. Such attacks effectively lead to denial-of-service conditions. It is important to note at this point that the smart grid exhibits a number of real time requirements, ranging from a delay constraint of 3 ms for trip protection in substations to a magnitude of minutes \cite{NIST2014,Laverty2010,Baigent2004}; therefore, under the premise that a real time system's response to a stimulus is considered correct and valid only if the response is given in a designated time frame (the \textit{real-time constraint}), any attack that may cause delays beyond the time constraints is also considered an attack on the system's availability.

\end{itemize}

\subsection{Attacks on confidentiality}
\label{subsec:attacksOnConfidentiality}
Data confidentiality can be threatened using many methods. A prominent method of data interception is to perform eavesdropping on wireless communications, which are used in the context of many smart grid implementations. \cite{ENISA2013} reports that electromagnetic radiation emitted by electronic components, such as CPU, display and keyboard can be used to infer information, however such techniques are less easily applicable and require elevated skills. Notably, even when the actual payload of communications cannot be captured or exploited (e.g. due to the fact that payloads are encrypted), communication patterns can be recorded -either through packet capture or through examination of emitted radiation- and subsequently analyzed to infer information such as device state changes from off to on.

Besides passive eavesdropping, active attacks against confidentiality are also possible. Man-in-the-middle attacks \cite{Conti2016} and spoofing attacks \cite{ENISA2013} can be launched by threat agents to intercept communications between smart grid nodes. Both types of attacks can be used to expose data from encrypted communication channels where peer node authentication is weak.

In addition to the targeting of communication channels, data can be stolen from nodes within the smart grid. In this type of attack, threat agents target the firmware or software of smart grid elements, such as advanced metering infrastructure (smart meters; concentrators) or even utility infrastructure, collecting data from multiple concentrators. Utility infrastructure is expected to be better protected, hence tampering with its firmware/software will be a harder task to achieve, however the expected payoff for attackers is significantly higher. If the firmware/software of the smart grid element is compromised, all data passing through the specific element can be copied to the threat agent. The same effect can be accomplished by seizing control of the nodes or specific services running on the nodes using alternate methods, including exploitation of default or weak passwords, buffer overflows and so forth.

\subsection{Attacks on integrity}
\label{subsec:attacksOnIntegrity}
In the context of a smart grid system attackers may employ different means to attack data integrity. Injection of false data \cite{Wang2013,Huang2013} relates to the introduction of counterfeit packets providing bad data to the utility company. False data injection is further subdivided into two categories (a) injection of carefully crafted data packets where measurements reported within the payload follow (to some extent) the laws of physics and (b) injection of packets whose payload does not comply with the laws of physics. The first category is more elaborate and may better evade detection, however it places some restrictions on the (false) values that may be reported. The load redistribution attack \cite{Yuan2011,Yuan2012} is a special type of attack on smart grids, where the attacker attempts to provide false measurements in a way that the overall energy consumption is maintained intact, however some nodes appear with under-reported consumption, at the expense of other nodes whose consumption is over-reported. The load redistribution attack may require full smart grid topology knowledge, however more elaborate algorithms allow for launching the attack when having only partial knowledge on the smart grid topology \cite{Liu2014}.

In order to inject false data, an attacker may employ different techniques: if communication channels are not authenticated or data is not securely transmitted, counterfeit data packets may be inserted into the data stream directly. Replay attacks \cite{ENISA2013} can be used to re-send the same packets multiple times, leading the utility company to compute erroneous consumption/demand data. Man-in-the-middle attacks \cite{Conti2016} allow attackers to intercept data from one communication endpoint, and subsequently analyze and falsify it, and finally send the falsified data to the originally intended recipient. Similarly to the case of attacks on confidentiality, besides falsification of data whilst it is transferred on the network, data can be falsified when it is processed or stored on the smart grid nodes: again, attacks against the firmware/software of nodes as well as attacks that exploit poor authentication/configuration on the nodes and the services they host can be used to obtain full or partial control of the smart grid nodes and subsequently inject false data. 

Another attack on data integrity in the context of smart grids is the so-called \textit{time modification of the gateway} \cite{ENISA2013} or, more generally, the \textit{time synchronization attack} \cite{Gunduz2020}. In this type of attacks, attackers aim at changing the relation between date / time measured consumption or production values in the meter data records \cite{ENISA2013}. Time synchronization attacks may also adversely impact the accuracy and effectiveness of a number of smart grid functions, such as event localization, monitoring voltage stability and fault detection on transmission lines \cite{Gunduz2020}. \cite{gong2012time} presents how GPS signals (which are used as a time source) are spoofed in order to realize time synchronization attacks.

\subsection{Attacks on availability}
\label{subsec:attacksOnAvailability}
Wireless channels used in smart grid networks, constitute a prominent target for attacks on availability, in order to hinder data flow. Application-layer denial-of-service (DoS) attacks, aiming to deplete host resources and disrupt the reception or processing of otherwise properly transmitted data are also possible.

\cite{Wang2013b} presents a taxonomy of known DoS attacks against smart grids, according to the OSI layer targeted by the attacks.
\begin{itemize}
\item \textit{Physical layer attacks}: This attack type typically jams the physical layer completely, disallowing data transmission between nodes connected via the affected channel. Jamming attacks are of low complexity for attackers, requiring elementary skill levels and commodity, inexpensive hardware, since only emission of signals within a designated spectrum range is necessary to achieve physical channel jamming. When the physical channel is jammed, no communication between nodes interconnected via the specific channel can be achieved.

\item \textit{MAC layer attacks}: The MAC layer is responsible for assigning unique addresses to the devices communicating via the network medium, while it additionally addresses arranges for applying rules for managing access to the common medium. Both these aspects may be attacked: malicious nodes may spoof other nodes' MAC addresses, and spoofing in combination with broadcasting of forged address resolution protocol (ARP) packets can be used to terminate the connections of all Intelligent Electronic Devices to the gateway node of a substation \cite{Wang2013b}. Additionally, malicious nodes may violate medium access rules to achieve better terms of service for themselves, degrading thus the quality of service offered to other nodes and increasing the probability that some real-time deadlines are violated; \cite{Cagalj2005} describes cases where malicious nodes use modified backoff parameters to re-transmit their packets sooner, after a collision has been detected on the shared medium.

\item \textit {Network and transport layer attacks}: Flooding is the most prevalent attack for this layer; \cite{Premaratne2015} presents an attack where ICMP packets generating a data flow of 10Mbps achieve to completely hinder communication with the victim node, whereas a data flow of 5Mbps degrade performance substantially. \cite{Choi2012} presents a SYN attack, where network/transport layer resources are depleted and no additional connections can be served, as well as a network/transport layer buffer overflow attack, leading to DoS. A buffer overflow attack, exploiting the vulnerabilities of actual smart grid devices, is also presented in \cite{Jin2011}.
\end{itemize}
While in theory attacks on the smart grid application layer to demote availability can be launched, the current literature does not report specific DoS application-layer attacks that have been launched against smart grid environments. This may be owing to the fact that the applications running on the public part of the smart grid network necessitate little processing and computing resources, and are thus less prone to be exploited in the context of DoS attacks. Complex smart grid-related applications, such as data mining and prediction software, run in private, well-protected environments, and -besides utilizing data generated from the public smart grid network- are completely isolated from it, and cannot therefore be directly targeted by DoS attacks.

\section{Attack detection and mitigation}
\label{sec:attackDetectionAndMitigation}

Some of the attacks against smart grids presented in section \ref{sec:attackTypes} are specific or tailored executions of generic attack patterns that can be launched against any system, while some other attacks are particular to the smart grid environment. Correspondingly, the first class of attacks may be detected and mitigated by applying well known methods from the systems security domain (which may be adequately modified or tailored to suite the smart grid context and/or operate optimally in it), while for the second class of attacks, new domain-specific methods must be developed.

\subsection{Detecting and mitigating attacks on confidentiality}
As noted in section \ref{subsec:attacksOnConfidentiality}, some attacks on confidentiality are passive, in the sense that they do not alter the data flows or the relevant nodes in any way, while others are active. Passive attacks are effectively based on eavesdropping, and therefore have no easily observable traits being very hard to detect. However, they can prominently be mitigated through the use of cryptographic techniques to protect the data in transit. For attacks based on traffic flow analysis, "filler packets" with no useful payload can be transmitted to disallow discerning between activity and tranquility periods. Finally, while exploitation of electromagnetic radiation emitted by electronic components is theoretically possible to infer activity or other information, the $complexity/utility$ ratio of these attacks is high, hence the probability of their appearance is correspondingly low. Notably, many components in the smart grid environment are restricted in CPU, memory and power capabilities, hence the use of lightweight encryption mechanisms, providing an appropriate balance between the achieved security level and resource consumption should be pursued.

Man-in-the-middle attacks fall in the active attacks category; these attacks may be detected by placing cryptographic signatures on the network packets and appropriately examining them. For this detection and mitigation path be effective, strong authentication mechanisms should be applied. 

Finally, in order to detect tampering with smart grid devices that can be used to steal data while it is being processed, the integrity of the devices' firmware, software and configuration can be validated. Device attestation \cite{Asokan2015} can be used to this effect, albeit device attestation protocols are complex in nature. Outgoing traffic from devices can also be monitored, to detect cases that data is communicated to unknown parties. Naturally, preventative measures can be applied, such as hardening of configurations, installing patches/firmware upgrades, changing default passwords, disabling unnecessary/bloatware services in the devices and so forth.

\subsection{Detecting and mitigating attacks on integrity}

False data injection is a major threat for smart grids; simplistic false data injection attacks can be detected by examining whether the reported data comply with the laws of physics, collating the data sourced from the smart grid's data channel with the measurements sourced from the smart grid's energy channel \cite{Drayer2018}. \cite{Singh2018} proposes a model taking into account more features of the power channel and exhibiting increased potential to detect false data injection attacks, even when realistic data are injected. Load redistribution attacks, can be detected and mitigated using different approaches, such as support vector models \cite{chu2020detecting}, machine learning \cite{Pinceti2018} or game theory \cite{Xiang2017}.

Replay attacks can be detected by providing sequence number in the packet payload and guarding against arbitrary sequence number modifications using techniques against sequence number attacks \cite{Gont2012} and/or cryptographic protection of sequence numbers or whole packets. Man-in-the-middle attacks can be confronted using strong authentication techniques, as discussed in subsection \ref{subsec:attacksOnConfidentiality}. Finally, \cite{Zhang2013} reports that time synchronization attacks can be detected and tackled by exploiting information from multiple layers (physical layer of the time synchronized measuring devices and the whole grid level); additionally, the physical infrastructure of the grid can be modelled as a dynamic on which system stability analysis will be applied to detect time synchronization attacks.

\subsection{Detecting and mitigating attacks on availability}

Attacks on the availability of the smart grid mostly observe the typical attack patterns in generic systems (network jamming; mac spoofing and selfish behaviour; network stack resource depletion); in this sense, standard DoS techniques employed in standard systems, such as mitigation of SYN flood attacks \cite{Kumar2018} or detection of excessive connection or request rates can be  applied in the smart grid context. Moreover, some special smart-grid specific detection and mitigation methods are proposed in the literature: \cite{Liu2017} presents an approach to offer increased resilience against jamming and enabling self-healing. \cite{Chlela2018} focuses on the microgrid domain, presenting an approach for confronting DoS attacks therein. Finally, \cite{Afianti2019} introduces multiuser dynamic cipher puzzle (M-DCP) equipped with TinySet, which is able to offer guaranteed confidentiality in the multiuser WSN authentication and lightweight DoS resistance.

\section{Conclusions}
\label{sec:conclusions}

In this paper we have presented a comprehensive overview of the threat landscape in the context of smart grids. These threats target the assets within smart grid installations and -if they are successfully realized- will demote their values. Since the effect of some compromises may be very severe, it is imperative that security grid managers are aware of the extent and severity of these consequences, while security officers in the domain must be knowledgeable about specific techniques exploited to realize attacks, as well as traits, methods and tools that can be exploited to detect and mitigate these attacks. Since the mitigation plan budget is typically constraint, the formulation of the mitigation plan should firstly review the key security goals of the particular smart grid deployment, and subsequently select and prioritize mitigations to meet the security budget, while minimizing the residual risk. To this end, this paper also offers a comprehensive list of key security goals related to the smart grid environment.

%

\section{acknowledgement}

\setlength{\columnsep}{14pt}%
\begin{wrapfigure}[3]{l}{.9cm}
\raisebox{-10pt}[0pt][0pt]{\includegraphics[height=.8cm]{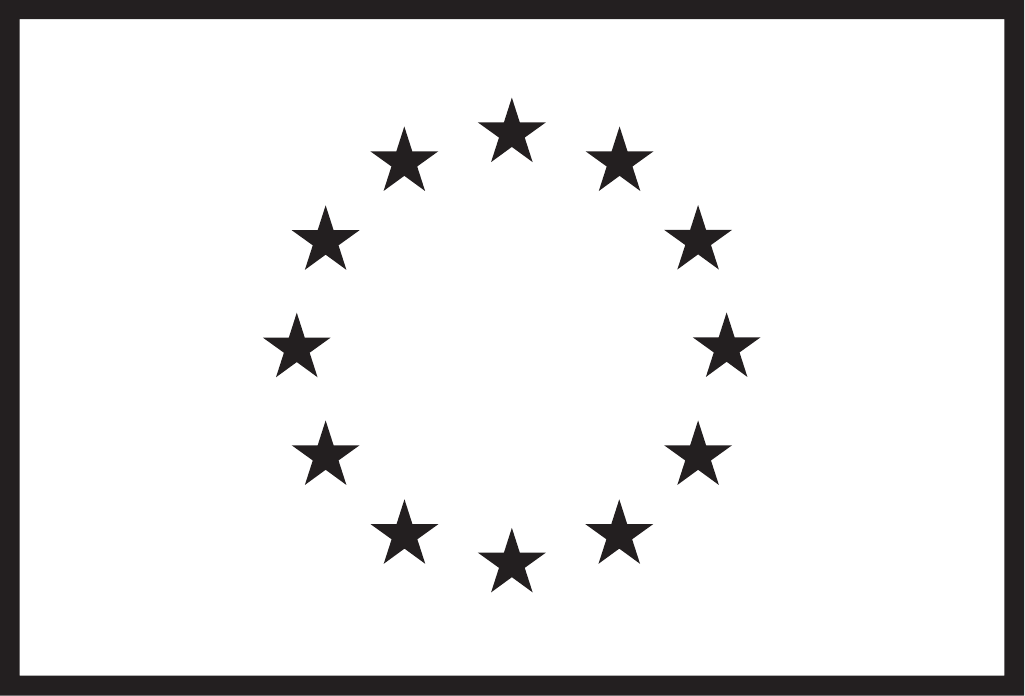}}%
\end{wrapfigure}%
This project has received funding from the European Union’s Horizon 2020 research and innovation programme under grant agreements no. 786698 and 833673. The work reflects only the authors’ view and the Agency is not responsible for any use that may be made of the information it contains.

\bibliography{Paper}
\end{document}